# Heterogeneous and rate-dependent streptavidin-biotin unbinding revealed by high-speed force spectroscopy and atomistic simulations

Felix Rico[1‡]*, Andreas Russek[2‡], Laura González[3], Helmut Grubmüller[2]*, and Simon Scheuring[4,5]*

[1] LAI U1067, Aix Marseille Univ, INSERM, CNRS, 163 avenue de Luminy, 13009 Marseille, France

[2] Department of Theoretical and Computational Biophysics, Am Faßberg 11, Göttingen, Germany

[3] Bioelectronics Group, Department of Electronics, Universitat de Barcelona, c/ Marti Franques 1, 08028 Barcelona, Spain

[4] Department of Anesthesiology, Weill Cornell Medical College, 1300 York Ave, New York, NY 10065, USA

[5] Departments of Physiology and Biophysics, Weill Cornell Medical College, 1300 York Ave, New York, NY 10065, USA

* Corresponding authors
‡ Equal contribution

## Abstract

Receptor-ligand interactions are essential for biological function and their binding strength is commonly explained in terms of static lock-and-key models based on molecular complementarity. However, detailed information of the full unbinding pathway is often lacking due, in part, to the static nature of atomic structures and ensemble averaging inherent to bulk biophysics approaches. Here we combine molecular dynamics and high-speed force spectroscopy on the streptavidin-biotin complex to determine the binding strength and unbinding pathways over the widest dynamic range. Experiment and simulation show excellent agreement at overlapping velocities and provided evidence of the unbinding mechanisms. During unbinding, biotin crosses multiple energy barriers and visits various intermediate states far from the binding pocket while streptavidin undergoes transient induced fits, all varying with loading rate. This multistate process slows down the transition to the unbound state and favors rebinding, thus explaining the long lifetime of the complex. We provide an atomistic, dynamic picture of the unbinding process, replacing a simple two-state picture with one that involves many routes to the lock and rate-dependent induced-fit motions for intermediates, which might be relevant for other receptor-ligand bonds.

## Highlights

Protein-ligand interactions are commonly described in terms of a two-state or a lock-and-key mechanism. To provide a more detailed and dynamic description of receptor-ligand bonds and their (un)binding path, we combined high-speed force spectroscopy and molecular dynamics simulations to probe the prototypical streptavidin-biotin complex. The excellent agreement observed, never used for force-field refinement, provide the most direct test of the 'computational microscope'. The so far largest dynamic range of loading rates explored (11 decades) enabled accurate reconstruction of the free-energy landscape. We revealed velocity-dependent unbinding pathways and intermediate states that enhance rebinding, explaining the long lifetime of the bond. We expect similar behavior in most receptor-ligand complexes, implying unbinding pathways governed by transient, timescale-dependent induced fits.

**Text**

\body Receptor/ligand bonds are at the core of almost every biological process. The early lock-and-key model including possible conformational changes of the binding partners is commonly accepted to describe the affinities and kinetic rates of receptor/ligand complexes, and is mainly based on molecular complementarity pictures from static structural data (1, 2). Over the past decades, an impressive amount of knowledge has been accumulated on the structural and energetic determinants of bound states, thus enabling the increasingly successful rational design of nano-molar binders for therapy as well as the quantitative prediction of binding processes and free energies from atomistic simulations (3). While protein folding and unfolding are thought to follow a multiplicity of pathways, the very mechanism of binding or unbinding of receptor/ligand complexes remains less investigated and is generally described by a simple two-state model, or by the lock-and-key analogy. Moreover, little is known on how the (un)binding dynamics is governed by the underlying microscopic processes — despite being key to a quantitative understanding of receptor-ligand complexes. Progress is mostly hampered by the lack of structural and thermodynamic information of the transient ligand/receptor conformations during unbinding, even for extensively studied systems such the complex formed by streptavidin (SA) and the small molecule biotin (b, vitamin H), one of the strongest non-covalent bonds known in nature.

Streptavidin forms the biotin-binding pocket with an eight-stranded, antiparallel beta-barrel capped by loop 3-4 (Fig. 1a). In the native, tetrameric SA form, loop 7-8 from an adjacent monomer provides a closing lid to the pocket (4). Biotin binds by forming an intricate and extensive network of hydrogen bonds with polar residues of SA (4, 5). Its high affinity ($K_D \sim 10^{-13}$ M) and long lifetime ($\tau \sim 10$ days, $k_{off} \sim 10^{-6}$ s$^{-1}$) (6-9), makes the SA/b system extensively used in biotechnology and biophysics. Dynamic forced disruption of the streptavidin-biotin complex by atomic force microscopy (AFM) and other techniques pioneered single molecule biomechanics (10-13) and provided estimates of the distance $x_\beta$ to the unbinding transition state and the intrinsic bond lifetime (13-15). Despite its seeming simplicity, AFM (11, 16-19), optical tweezers (20), and biomembrane force probe (10, 21) experiments of streptavidin/biotin unbinding have reported dissimilar results suggesting an impressive complexity and heterogeneity in the unbinding pathway (see SI text). Furthermore, the results from single molecule studies were incompatible with ensemble bond-lifetime measurements (6).

Although recent experimental developments accessing the microsecond timescale have shed light into the complexity of single molecule transition paths of protein and nucleic acid (un)folding (22-25), the amount of transient structural information extracted from single molecule experiments is rather limited. Therefore, most structural knowledge on unbinding/unfolding processes has been derived from atomistic simulations that were often limited to short time-scales inaccessible to experiments and therefore not rigorously validated (24-28). Thus today, a direct relationship between the energy landscape and the dynamic structural details of these seemingly simple biomolecular processes is missing. As a result, it is still unclear (a) how biotin precisely outlives days and unbinds under load, (b) how and where biotin is located at the point of rupture and how the respective intermediates are stabilized, (c) if there is only one or possibly several unbinding pathways, and if so, (d) to what extent the unbinding paths change with loading rate. Here we address these questions by combining high-speed force spectroscopy (HS-FS) and fully atomistic simulations to observe biotin unbinding from streptavidin over eleven decades of loading rates. We show that the unbinding pathway of the small molecule biotin from SA is much more complex than a "key that leaves a lock" and reveal a multitude of pathways and intermediate binding sites far from the binding pocket, similar to the various pathways and intermediate states of protein (un)folding.

For the HS-FS experiments we used microcantilevers functionalized with biotin to probe the force required to rupture individual SA-b bonds at various loading rates (**Fig. 1a, left**). The use of microcantilevers with response time of ~0.5 µs and reading out the reflected laser beam at 0.05µs (high sampling rates up to 20 MSamples/s) allowed tracking the cantilever position while pulling at velocities up to ~30,000 µm/s, almost an order of magnitude faster than previous HS-FS measurements and about 1000 times faster than conventional AFM FS measurements (11, 19, 29). All atom steered molecular dynamics (SMD) simulations precisely mimicked the experimental setup by using the fully solvated tetrameric structure of streptavidin (PDB 3RY2 (4)) and by pulling biotin using two springs in series, using the worm like chain model describing the PEG-linker and a linear spring for the cantilever, whose end was moved at constant velocity (**Fig. 1a, right, and SI movies 1-3**). The overall applied pulling velocities ranged from 0.05 µm/s to 30,000 µm/s in HS-FS experiments (**Fig. 1b**), and from 1,000 µm/s to $5 \cdot 10^{10}$ µm/s in SMD simulations (**Fig. 1c**), resulting in a combined range of loading rates from ~100 pN/s to ~$10^{13}$ pN/s, covering eleven decades. Importantly, the wide experimental dynamic range up to such fast rates, together with recent simulation advances (30, 31), allowed direct overlap and comparison with *in silico* simulation loading rates over an entire decade between $10^8$-$10^9$ pN/s (**Fig. 2 and SI**).

The measured force curves showed a characteristic curve of increasing force due to stretching of the flexible PEG linker (**Fig. 1**), signature of the specificity of the interaction (32) (cf. Materials and Methods). **Figure 2** shows the dynamic force spectrum obtained both from the most probable rupture forces at each loading rate (in pN/s) in the experiments (circles) as well as from the average rupture force of 10-20 simulations per loading rate (triangles). At the overlapping loading rates, rupture forces from experiments and simulations agreed very well, thereby providing independent validation for the MD simulations. At the lowest loading rates, from ~$10^2$ pN/s up to ~$10^6$ pN/s (~$10^3$ µm/s pulling velocity), rupture forces increased almost linearly with the logarithm of the loading rate, indicating one single dominant barrier in this loading rate regime. At faster loading rates, steeper slopes are observed. This behavior has been interpreted before in several ways: (a) multiple transition barriers along the one-dimensional energy landscape, inner barriers becoming dominant at high loading rates and outer barriers, at low rates (10, 19); (b) force-induced shortening of the distance to the transition state (14, 33), and (c) a transition from a thermally activated to a so-called deterministic regime (15, 24, 34). Another possible cause (d) might be that the cantilever response affects the force spectrum for pulling timescales shorter than the cantilever response time (35, 36). Actually, previous experiments using devices with a dynamic response slower than HS-FS cantilevers (response time $\tau_c$~50-500 µs, effective diffusion constant $D_b$~$10^2$-$10^3$ nm$^2$/s, compared to $\tau_c$~0.5 µs, $D_b$~$10^5$ nm$^2$/s) have reported a marked slope increase at loading rates ~$10^4$ pN/s (10, 16, 19, 29), while our first slope increase occurred at ~$10^6$ pN/s. This suggests that this possible effect (d) is reduced using HS-FS microcantilevers, which allows orders of magnitude faster loading rates before this possible effect may appear.

By virtue of the broad range of loading rates covered here, the combined dynamic force spectrum contains more information on the free energy landscape of unbinding than has been accessible before. As detailed below, all three possible explanations (a-c) seem to contribute to the shape of the energy landscape. In particular, single barrier models did not describe the entire dynamic force spectrum satisfactorily, supporting the presence of a more complex energy landscape with multiple barriers (14, 15, 34, 36, 37) (**SI Fig. S7**). To avoid approximations inherent to analytic theories, we instead performed Brownian dynamics simulations using a more complex energy landscape with two barriers and varied the shape and height of these barriers (inset in **Fig. 2**, see also SI) until the best agreement with the dynamic force spectrum (blue line in **Fig. 2**) was obtained. Importantly, this free energy landscape explains both experiment and simulation over the whole 11 decades of loading rates. The energy landscape has a first (inner) ~17 $k_B$T barrier at 0.19 nm, which determines the force spectrum slope at loading rates faster than $10^6$ pN/s, and a second ~21 $k_B$T unbinding barrier further out at 0.44 nm (**Fig. 2** inset, blue line). The longer rupture length of the second barrier, which becomes rate-limiting only at lower unbinding rates, gives rise to the shallower slope at loading rates below $10^6$ pN/s. We note that the MD simulations suggest intermediate states (i.e., a well between these two barriers, sketched as a dashed blue line in the inset), which however cannot be extracted (nor ruled out) by analysing the force spectrum alone (33). Also, a third unbinding barrier further out at distances larger than 0.44 nm (indicated by the dashed red line in the inset) is seen in our experimental force curves and MD simulations, and will be discussed below. However, in the force spectrum this barrier would only become kinetically dominant at lower loading rates than probed our HS-FS experiments and is therefore not seen.

According to the Bell-Evans theory, one would expect two distinct slopes in the force spectrum, corresponding to the two kinetically relevant unbinding barriers(13). In our dynamic force spectrum, we observe however a rather continuous slope increase up to $10^{11}$ pN/s attributed to a force-induced shortening of the rupture length of these two barriers, as predicted by theories that take the shape of the barriers into account (14, 36). Finally, the slight deviation of the BD curve at very fast loading rates >$10^{11}$ pN/s may indicate a transition from a diffusion-dominated (Bell-Evans regime) to a deterministic regime (15, 34). For SA-b, this critical loading rate ($\dot{F}_c$>>$F_c D x_\beta^{-2}$) ~$10^{11}$ pN/s (or ~$10^{10}$ nm/s) is orders of magnitude faster than that observed in previous HS-FS experiments of titin unfolding (~$10^7$ pN/nm, ~$10^6$ nm/s) (24). This was expected since the transition is supposed to emerge when the pulling rate is faster than the intrinsic time required for the complex to explore its energy landscape. This intrinsic time was ($x_\beta^2/D$) ~0.2 ms for titin I91, but much shorter (~1 ns) for SA/b, which is reasonable given the less pronounced structural changes of the biotin molecule compared to partial titin domain unfolding.

The MD unbinding simulations (>300 in total) provided structural information on the loading rate dependent unbinding paths. **Figure 3A** shows the distribution of center of mass (COM) distances from all MD trajectories between the SA binding pocket, defined as the set of amino acids that interact with the bound biotin in the static crystal structure, and the biotin molecule. The peak at 0 nm represents the bound state followed by two consecutive minima at ~0.25 nm and ~0.5 nm. These values are slightly left of the positions of the two barriers (~0.19 nm and ~0.44 nm) obtained from the BD fit of the force spectrum and suggest the COM distance as an appropriate reaction coordinate of unbinding. Interestingly, the peak between and to the right of these two minima suggests that, at least upon force application, one or several metastable states appear, as a result of the tilted energy landscape. Importantly these metastable states will favor rebinding at sufficiently slow pulling rates (37, 38). The simulations show that most of the H-bonds between biotin and the SA binding pocket remain intact until biotin has moved ~0.15 nm towards the outside of the SA binding pocket. At the distance corresponding to the first barrier, the H-bonds between residues Ser27, Tyr43, Asn49, and Asp128 and the biotin, rather parallel to the pulling direction, rupture. Escape from the binding pocket occurs only after the second barrier at ~0.5 nm, where most of the remaining H-bonds between biotin and SA (mainly with residues Asn49, Tyr54 and Arg84) rupture (**Fig. 4B and SI Fig. S9**). Notably, these H-bonds are nearly perpendicular to the unbinding direction (long axis of the binding pocket), which implies a shear force, and only simultaneous failure of all H-bonds lead to dissociation, with subsequent transient formation of a different H-bond network (**SI Fig. S9**). A similar mechanism has been observed before as key to providing stability against forced protein unfolding (39, 40). Overall, the geometry of rupture and reconfiguration of the H-bond network between biotin and adjacent amino acids of the binding pocket seems to represent the main determinants of the dynamic force spectrum in **Fig. 2**, described by the energy landscape with two barriers.

One might assume that linear extrapolation of the force spectrum to zero force should yield a time scale similar to the spontaneous SA-b unbinding off-rate $k_{off}$ ~$10^{-6}$ s$^{-1}$ obtained from bulk equilibrium experiments (7). However, as in previous single molecule force experiments (10, 19) (see SI), **Figure 2** would suggest a much faster off-rate of ~1 s$^{-1}$. Notably, also the 21 $k_B T$ barrier (blue line in the inset of **Fig. 2**) in the unbinding energy landscape is ~19 $k_B T$ lower than the calorimetric SA-b binding free energy of ~40 $k_B T$ (7). Further, a Kramers estimate using an attempt frequency of $10^9$ s$^{-1}$ would also predict a ~1 s$^{-1}$ off-rate. Whereas the end states of enforced and spontaneous unbinding are not the same and, hence, the respective (un)binding free energy differences are not expected to fully agree, such a large discrepancy is unexpected. To reconcile forced and equilibrium unbinding, a third barrier located further out on the unbinding pathway should be present (as depicted in **Fig. 2** inset, red dashed line), as was speculated before from indirect evidence (10, 21). Such a barrier would show up in a dynamic force spectrum at loading rates much lower than accessible to experiments (and certainly to simulations) and, therefore, remained so far unobserved. One would, however, expect to observe interactions between SA and biotin farther out of the binding pocket, as structural determinants of this barrier.

Indeed, our MD simulations revealed such interactions and corresponding intermediate states. In particular, the distribution of COM distances between the SA binding pocket and the biotin molecule displays a pronounced peak after the second minimum followed by smaller peaks at distances up to 1.5 nm (**Fig. 3A and SI**). Detailed inspection of the individual MD trajectories showed as many as eight transient H-bonds (mainly to ASN49, GLU51, and TYR54) formed with biotin ~1 nm away from the binding pocket (**Fig. 3A**). Importantly, the force profiles (**Fig. 3B**) from these trajectories (~15% of all trajectories) displayed adhesive interactions after and at a lower value than the main force peak and, therefore, these states are not seen in the dynamic force spectrum. In these events, the force applied to biotin displays a drop due to the exit from the binding pocket and a subsequent intermediate force plateau with a final drop due to complete detachment. At the slowest MD velocities, this transient unbinding state lasted up to several hundred nanoseconds, such that it should also be detectable using HS-AFM microcantilevers with sub-µs resolution.

To test this hypothesis, we analyzed in further detail the individual experimental force curves. Remarkably, we observed a similar signature in about 5% of the HS-FS unbinding events with a transient force plateau during the snap off of the cantilever (**Fig. 3C** and **SI Figs. S5** and **S6**). These transient, µs-long events were observed over the full range of experimental loading rates and provide an experimental signature of the transient outer states. The low occurrence of these events in HS-FS curves may be due to their short lifetime. Moreover, the distance from the force peak to the transient binding had an average value ~1 nm extending up to 3 nm (**Fig. 3D**), similar to the distances seen in the MD trajectories (**Fig. 3A**). Therefore, the combination of HS-FS and MD simulations of SA/b forced unbinding provided first direct experimental evidence and a structural description of an outer binding state that may be at the basis of the SA/b sturdiness.

The large number of experiments and simulations allowed us to characterize the average lifetime $\tau$ of these outer binding states. **Figure 3E** shows that $\tau$ ranges from 0.001 μs to 100 μs for forces $F_i$ between 500 pN down to 20 pN (including MD and HS-FS data). Hence, excellent agreement between experiments and simulations is seen also in the time domain. The average lifetime decayed exponentially with force and can be described by a single barrier (14) of 12 $k_BT$ additional height with a rupture length of 0.16 nm and, notably, of ~16 μs-lifetime at zero force (**Fig. 3D**). This third barrier (red dashed line in **Fig. 2**, inset) is located outside the second barrier (blue line), at a distance of 0.60 nm from the bound state (red line in Fig. 2 inset), and adds a further step upwards to the energy landscape towards the fully unbound state. This distance correlates with a minimum in the COM distance at ~0.7 nm. Although the exact minimum preceding this outer barrier is difficult to pinpoint, metastable states before this barrier are expected upon force application, as reflected from the peak in the COM distribution at ~0.6. This barrier further slows down biotin unbinding by several orders of magnitude, thereby reconciling it with the observed slow equilibrium off-rate. The respective interactions between biotin and the outside of the binding pocket should favor rebinding events, in particular at low and zero forces. Although rare, back-and-forth fluctuations between intermediate states were actually observed in some of the MD trajectories (**Fig. 4**).

As shown in **Fig. 3D**, the experimental distance to the outer binding increased with the pulling velocity suggesting that shorter jumps occur more often at slow pulling. This suggests that, although effectively described by a single barrier, the outer barrier may involve not only one but several intermediate states, not directly resolved experimentally and with varying occupancies that depend on the pulling velocity. This notion is further supported by the various peaks observed in the COM distance outside of the binding pocket, which allowed characterizing the four most populated intermediate states ('Int 1-4', **Fig. 4A**). Importantly, various unbinding paths were seen in the trajectories. The large number of atomistic simulations for each loading rate allowed us, finally, to study to what extent the observed unbinding pathways change with loading rate. Under high load, mainly two intermediate binding states ('Int1' and 'Int2') are visited along the unbinding pathway. At lower loading rates, states 'Int3' and 'Int4', farther out, are also visited. Likely these, and even farther outside lying intermediates, provide a rugged funnel (41-43) for rebinding under equilibrium conditions.

Calculating the average energy of the H-bonds between biotin and individual amino acids in the binding pocket from all MD trajectories as a function of the biotin position (**Fig. 4B and S9**) allowed us to extract structural snapshots of each intermediate state (**Fig. 4C**). Whereas the inner intermediate states showed strongest interactions between the ureido moiety of biotin and residues Ser27, Tyr43, and Asp128, respectively, at later stages of unbinding other bonds, largely overlooked so far, become relevant, such as Arg84, Glu51 and Tyr54, at COM distances of up to 1.5 nm from the bound state (**Fig. 4B**). While streptavidin modifications of these three residues have reported lower biotin affinity (6, 44, 45), they have not been expected to be involved in biotin binding due to their large separation from the binding pocket and their little contribution to the bound state, underscoring the impact of features along the unbinding path on binding kinetics.

One might speculate that SA has evolved in tetrameric form because it allows for even larger binding affinity due to inter-monomeric stabilization, with the 7-8 loop providing the key inter-protomer interaction (5). To test this idea, we repeated our steered MD simulations using monomeric streptavidin – which is difficult experimentally. As suggested before from high loading rate simulations on avidin-biotin unbinding (27), rupture forces of the monomer were systematically 10-20% lower than for the tetramer over the whole loading rate range (**Fig. S8 and SI movies 4-6**), thus further supporting this hypothesis. Closer structural analysis of the unbinding paths suggests that these reduced unbinding forces are due to (a) lacking inter-monomeric interactions (specifically to the 7-8 loop of the adjacent protomer (27)) and (b) an increased heterogeneity of unbinding paths, the larger entropy of which further reduces the unbinding barrier.

Strikingly, the unbinding rate dependent heterogeneity and occupancies of intermediate states are accompanied by rate-dependent, non-equilibrium dynamics of the SA structure while biotin moves towards the unbound state. We observed induced-fit motions (of the binding pocket and adjacent loops) outside the fully bound state, along the (un)binding path. As an example, loop 3-4 switches between an open and a closed conformation during unbinding, depending on the distance between biotin and the binding pocket (**Fig S10**). These non-equilibrium conformational changes are more pronounced at slower loading rates, for which the loop has more time to fluctuate and equilibrate (**Fig S10**). Therefore, it should also occur in the AFM experiments as well as during spontaneous unbinding. This finding suggests that 'induced fit' and other non-equilibrium conformational changes of SA control not only the bound state, but also the transient energetics and kinetics along the binding and unbinding paths, and in a loading rate-dependent manner.

The combination of HS-FS and MD simulations at overlapping loading rates allowed us to obtain a dynamic and atomistic description of a receptor-ligand unbinding process. Characterizing enforced streptavidin/biotin unbinding over an unprecedentedly large range of loading rates enabled us to characterize large portions of the underlying energy landscape, which would not have been accessible by one of the two methods alone. Notably, it also allowed for a most direct comparison between AFM experiment and MD simulation. The excellent agreement of rupture forces at overlapping loading rates – an observable that has never been used for force field parameterization – underscores the predictive power of atomistic MD simulations.

Single barrier theoretical models successfully describe the spectrum over a wide range at high loading rates and serve to interpret our results (SI Fig. S7). The fitted BSK model predicts a transition towards the deterministic regime at rates $\sim10^{11}$ pN/s only reached by MD simulations, that may explain the change in slope at this loading rate (15). The recently developed CHS model, which described remarkably well the DFS, predicts instead a kinetically ductile regime described as gradual stretching and shortening of the distance to the transition state under force before unbinding, which helps understand the curvature observed at high loading rates (Figs. 2 and S7). The low value, close to zero, of the unitless kinetic brittleness ($\mu\sim10^{-6}$, SI Fig. S7 (36)) found for our spectrum is consistent with a ductile behavior for SA/b. Inclusion of the low loading rate regime requires multiple barriers (38), and the full dynamic force spectrum over 11 decades is only accurately described by BD simulations. Finally, outer barriers at distances beyond the energy landscape of Fig. 2 (inset, blue line) would enhance rebinding and likely emerge in the force spectrum at loading rates lower than the ones explored in our work (but were detected as transient binding events in both HS-FS and SMDS). Therefore, combinations of available analytical theories seem to be required to fully explain the force spectrum over the whole dynamic range, including a more refined description taking into account the dynamic nature of the energy landscape.

Our concerted approach revealed multiple unbinding pathways and nonequilibrium conformational changes of the SA binding pocket dependent on loading rate, with a detailed description of the pathway fluxes. In particular, outer intermediates were found that affect binding energetics and kinetics. Future combination of HS-FS and MD simulations will answer whether these proposed mechanisms found for SA/b are specific to this bond or – in our view more likely than not – a common feature of many receptor/ligand complexes and thus a general mechanism of regulating binding kinetics. If this were the case, the study of intermediates and their mechanism would be instrumental in improving the binding kinetics and specificity of drug-like compounds. Taken together, our results suggest that the current static picture based on the bound state may need to be extended in terms of many routes to (un)binding as well as multiple, transient and nonequilibrium induced fits, resembling a combination lock, where several intermediate positions have to be visited for final release. The development of theoretical models taking into account the dynamic nature of the energy landscape will help us better understanding biological bonds.

Figures

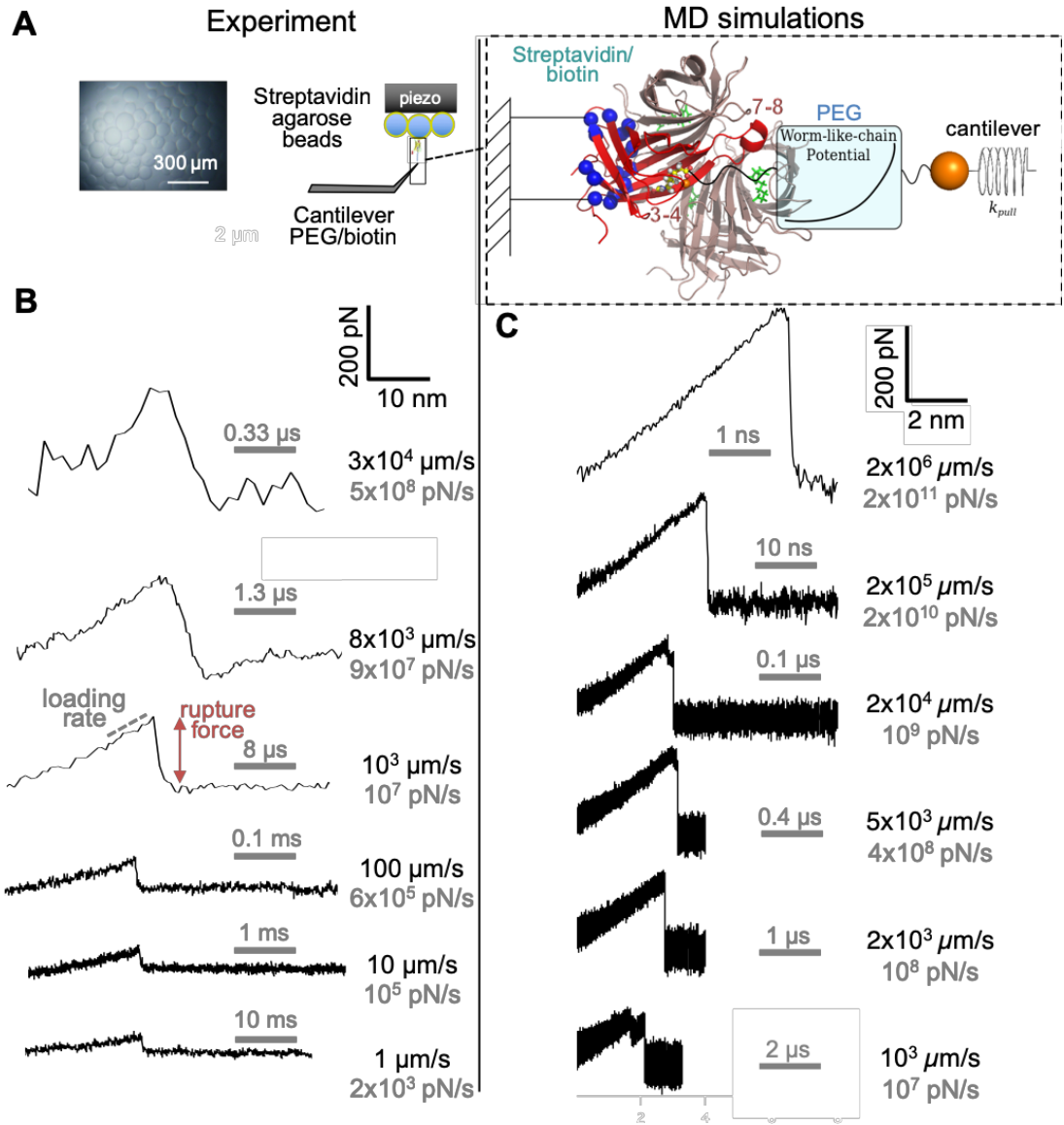

**Fig. 1 High-speed and molecular dynamics force spectroscopy of streptavidin-biotin unbinding.**

**(A)** Left: High-speed force spectroscopy (HS-FS) setup. Streptavidin agarose beads (top left) were immobilized on the sample surface while biotin was covalently attached to the microcantilever (bottom left) through a polyethylene glycol (PEG) linker (contour length ~10 nm). Right: MD simulations setup and the streptavidin-biotin tetramer used in the MD simulations with labeled loops 3-4 and 7-8 (PDB 3RY2). **(B)** Experimental force-distance traces at velocities from 1 μm/s to 30,556 μm/s revealing bond rupture events using AC7 (top curve) and AC10 cantilevers (bottom curves). The stretching profile of the experimental curves was the results of stretching the PEG linker and deforming the agarose bead to which streptavidin was linked. **(C)** Force-extension traces from the MD simulations at retraction velocities from 2000 μm/s to 2x10$^7$ μm/s.

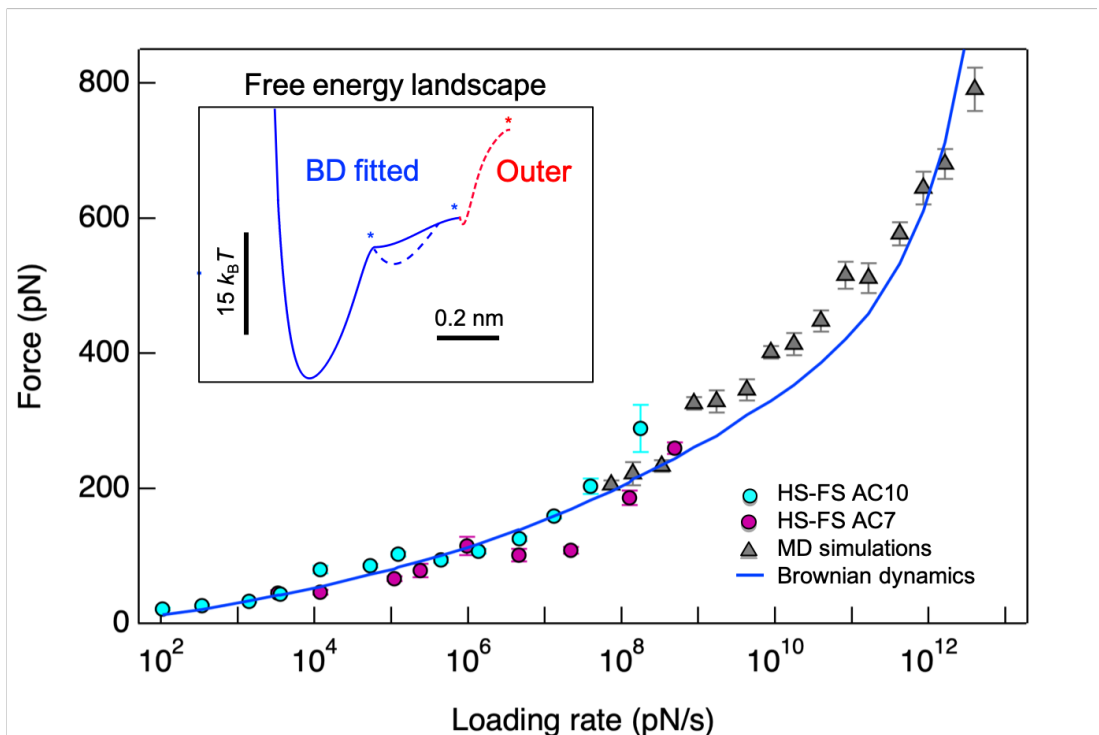

**Fig. 2. Dynamic force spectrum of streptavidin/biotin unbinding**. Most probable rupture forces (±SEM) from HS-FS experiments (●, using regular AC10 and fast AC7 cantilevers, cyan and magenta, respectively) and MD simulations (▲). The blue line represents a Brownian dynamics fit to the whole force spectrum. The inset shows the resulting equilibrium free energy landscape as a blue line, which revealed two transition barriers (with parameters $D$ = 4x10$^7$ nm$^2$/s, $\Delta G_1$ = 17 k$_B$T, $\Delta G_2$ = 21 k$_B$T, x$_{\beta 1}$ = 0.19 nm, and x$_{\beta 2}$ = 0.44 nm) and a possible intermediate state between the two (blue dashed line). The red dashed line sketches a third, outer barrier (see Fig. 3 and text), which cannot be extracted from the dynamic force spectrum.

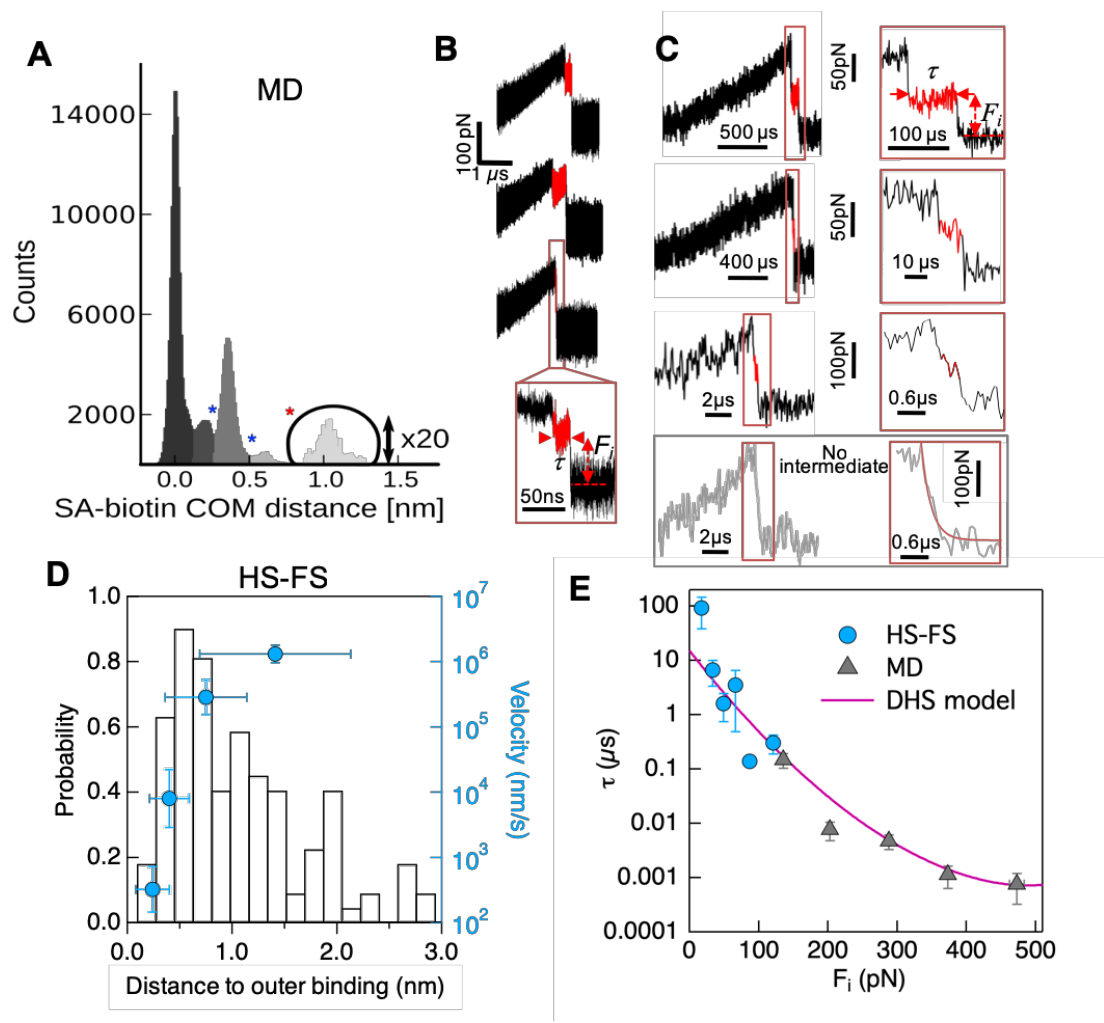

**Fig. 3. Outer barrier.** (A) Distribution of center of mass distances between streptavidin and biotin for all MD trajectories during unbinding. The first two minima coincide with the barrier positions obtained from the Brownian dynamics simulations fit of the force spectrum in the inset of Fig. 2. (B) Examples of MD and (C, top three) HS-FS force-time curves showing outer binding. The lifetime and applied force of outer binding events was determined as shown (red lines) and described in the Methods Section. The bottom curve shows an example for which no intermediate was observed. The red line shows an exponential fit with decay time ~0.28 μs. (D) Distribution of experimental distance to outer binding (left axis) and pulling velocity dependence of the average distance (blue, right axis) (E) Outer binding lifetime vs. applied force from HS-FS (●) and MD simulations (▲). The solid line shows the best fit to the force dependent lifetime DHS model (14) with parameters $\tau_0$=16±7 μs, $x_\beta$=0.16±0.10 nm and $\Delta G$=12±6 $k_B T$. This distance the outer barrier ($x_\beta$) should be added to the position of the second barrier at $x_{\beta 2}$ = 0.60 nm in Fig. 2 (sketched as a dashed red line in the inset).

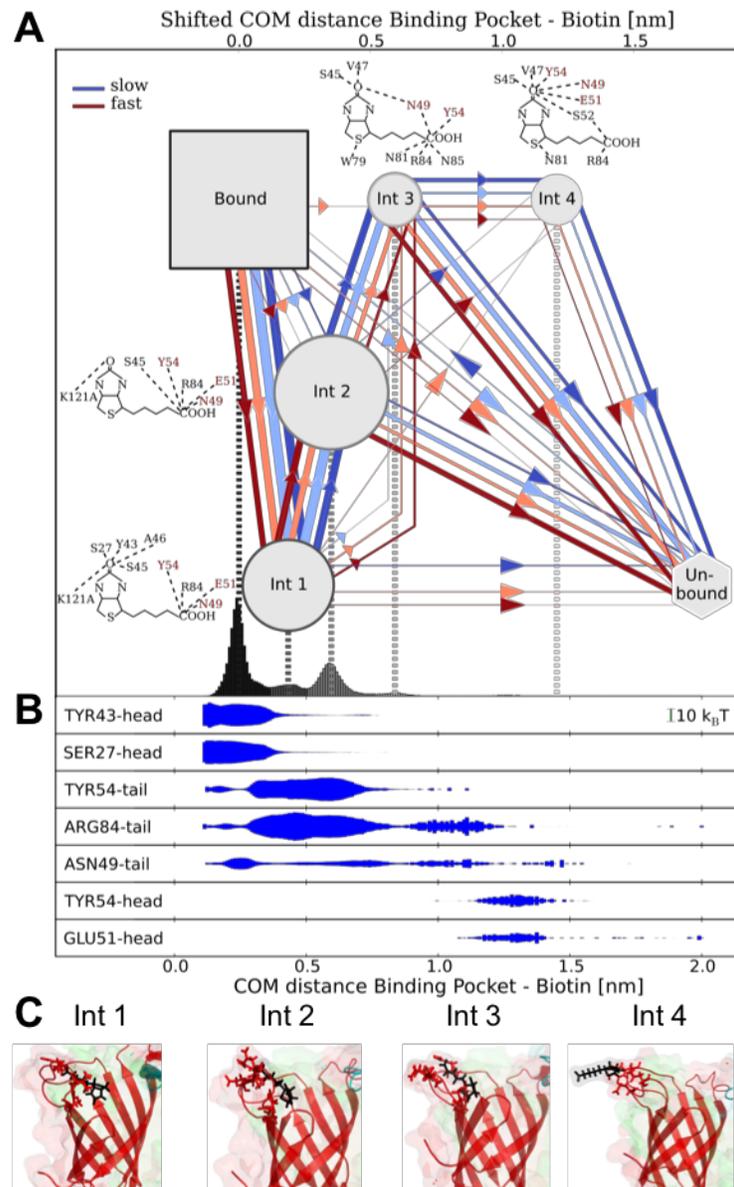

**Fig. 4. Dynamic multiplicity of streptavidin-biotin unbinding pathways.** (A) Bound and intermediate states with unbinding pathways observed during forced dissociation in MD trajectories. The line color reflects pulling velocity from slow (blue) to fast (red) and thickness reflects passage frequency. Biotin molecular representations show overlays of the hydrogen bonds with streptavidin residues for the four intermediate states. The amino acids with the strongest intermediate interactions are labeled in red. (B) The energy of the H-bonds between biotin and the most important residues is shown below as a function of the COM distance. (C) Structural snapshots of the different intermediate states showing H-bonds between SA residues and biotin.

/body

## Materials and Methods
### HS-AFM tips and sample preparation
Streptavidin coated 4% agarose beads (Sigma) were immobilized on the sample surface by embedding them in a thin agarose layer. Biotin was covalently attached to the cantilever through a polyethylene glycol (PEG) linker (stretched length ~10 nm, Fig. 1A). Briefly, HS-AFM (AC10DS and AC7) cantilevers (Olympus, Japan) were rinsed with acetone for 10 min, plasma cleaned for 5 min in $O_2$ and immersed in a solution of 10-20 mg/ml silane-PEG-biotin (1 kDa, Nanocs Inc, NY) in ethanol/water (95/5). After 2 h incubation, cantilevers are rinsed with ultrapure water and stored at 4ºC until use.

### HS-FS measurements
HS-FS measurements were carried out on an HS-AFM (RIBM, Japan) featuring a high-speed acquisition board system (PXI, National Instruments, Texas, USA) to control the z−movement and acquire force curves at sampling rates up to 20 MS/s. Two types of cantilevers with submicrosecond time response and small viscous damping were used: AC10 cantilevers with 600 kHz resonance frequency in liquid, 0.1 N/m spring constant, quality factor of 0.9, and 0.09 pN $\mu m^{-1}$ $s^{-1}$ viscous coefficient; and shorter AC7 cantilevers, with 1.3 MHz resonance frequency in liquid, 0.6 N/m spring constant, quality factor of 0.6, and 0.05 pN $\mu m^{-1}$ $s^{-1}$ viscous coefficient. The spring constant of the cantilevers was determined in air using the Sader method [46]. The optical lever sensitivity was then determined in liquid from the thermal spectrum and the known spring constant [24] [47]. Short HS-AFM cantilevers were placed on the cantilever holder immersed in the fluid cell with PBS (phosphate-buffered saline) buffer and placed on the HS-AFM. The streptavidin agarose functionalized sample-stage was then mounted onto the fluid cell. Force curves were collected approaching the sample at a constant speed of 10 μm/s and indenting the streptavidin agarose with the biotinylated tip for 0.5 s at a constant force of <500 pN. The sample was then retracted at varying retraction velocities from 0.010 μm/s to 30,000 μm/s. Free biotin was titrated into the fluid cell to achieve a lower binding frequency of ~5% favoring single-bond interactions, assuring that most of the events (>95%) reported single bond ruptures [48, 49]. The reduction in binding frequency after adding free biotin further confirmed the specificity of the interaction. Representative examples of retraction force curves are shown in figure 1B. The sampling rate was set between 2 MS/s and 20 MS/s depending on the retraction speed of the experiment.

### HS-FS data processing
Rupture events were detected using semi-automatic home-made routines (Matlab, Mathworks) from the numerical derivative of the force curves by defining a threshold established by the zero-force baseline noise and depending on the sampling frequency and pulling rate [24]. Force curves presenting rupture events were analyzed to measure the rupture forces, rupture lengths and effective spring constant (Fig. 1B). Loading rates were determined from the slope before rupture (2-3 nm) of each force versus time trace. Rupture forces were pooled by loading rate and the corresponding histograms generated (Supplementary Figs. S1 and S2). Most probable rupture forces and standard deviations of rupture forces at each loading rate were calculated from Gaussian fits (Supplementary Figs. S1 and S2). When two peaks were clearly distinguishable in the force histograms, distributions were fitted using a bimodal function imposing the center of the second peak to be 2 times that of the first peak. Thus, the first center peak value was used as the average rupture force (Supplementary Figs. S1 and S2).

### HS-FS transient binding
From the list of rupture events, transient or intermediate binding events were determined from the numerical derivative of the force drop region: from the main rupture event to the zero-force baseline (SI Figs. S5 and S6). The numerical derivative of the force drop region was calculated by the 5-point stencil method (50). An average filter with a window ranging from 0 to 25 µs was used according to pulling velocity and sampling rate. At the lowest velocities, due to the reduced sampling rate and the lower rupture forces, shortest events were likely missed. Transient binding events were defined from the time interval between the two most prominent minima within the force drop region including at least one data point of the force derivative with positive value. If this condition was not accomplished or the mean force value for the transient event was within one standard deviation of the zero-force baseline, no rebinding event was considered. The lifetime of the intermediate state was determined from the time interval between these two minima. The force level was deteremined as the average force of the interval. The distance to the intermediate state (Fig. 3D) was calculated from the force versus extension curves as the distance from the extension at which the main rupture occurs to the average extension of the interval.

**HS-FS viscous drag correction**

Viscous drag coefficients were calculated for each cantilever type from retraction curves by measuring the drag force exerted on the cantilever at various velocities near the substrate (<200 nm) divided by the retraction velocity (51, 52). The cantilever coefficients on agarose beads were 0.09 pN/(µm/s) and 0.05 pN/(µm/s) for AC10 and AC7 cantilevers, respectively. The viscous drag was corrected by multiplying the viscous drag coefficient by the relative tip velocity. This correction was only significant (~8% force correction) at the highest velocities achieved with each cantilever type.

**Molecular dynamics simulations**

Force probe molecular dynamics simulations were carried out using the Amber99sb (53) force field together with TIP3P (54) water model and virtual interaction sites (55). The electrostatic interaction was calculated with particle-mesh Ewald {Darden, 1993 #1067} with a real space cut-off of 1 nm, a grid spacing of 0.12 nm and cubic interpolation. The same cut-off length was used for the Van-der-Waals interaction. All atom bonds were constraint using LINCS algorithm (56). All simulations were performed at a constant number of particles and a constant temperature coupled to a heat bath at 300 K using the velocity rescaling method (57) and a coupling constant of 0.2 ps. We used the Verlet algorithm (58) with a time step of 4 fs to integrate the equation of motion. In the experimental setup, the cantilever, here described by a harmonic potential with a spring constant $k_{pull}$ = 100 pN/nm, was attached to Biotin via a PEG-linker. The harmonic potential was moved away from the binding pocket at 12 different velocities, ranging from 0.001 m/s up to 50 m/s. Depending on the applied loading rate the center of the harmonic pulling potential moved up to 12 nm until the simulation was stopped, ensuring that biotin is completely unbound from Streptavidin. In the simulations the linker was described by a worm like chain (WLC) potential

$$U_{wlc}(x) = \frac{k_B T l_c}{4 l_p} \left(1 - \frac{x + x_{start}}{l_c}\right)^{-1} - \frac{k_B T (x + x_{start})}{4 l_p} + \frac{k_B T (x + x_{start})^2}{2 l_c l_p}$$

with a persistence length $l_p$ = 1.4 nm and a contour length $l_c$= 10 nm similar to the experimental values (Fig. 1 sim setup), which was added to GROMACS as a tabulated bonded potential (54). To keep the simulation box size small we shifted the WLC potential by a constant $x_{start} = 7nm$, nevertheless making sure that the forces required to stretch the linker were lower than those to rupture the complex. All simulations were performed at a sodium chloride concentration of 150 mM. The simulation box sizes perpendicular to the pulling direction were chosen to be 8.5 nm, which is a minimal distance of 0.7 nm between the solute and the periodic boundaries, leading to a minimum distance between Streptavidin and its periodic image of more than 1.4 nm. The box length in pulling direction varied between 18 nm for the faster and 13 nm for the slowest simulations.

The starting structure for the simulations was based on the tetramer PDB-ID 3RY2 (4). In case of the tetramer, only one biotin was pulled out of the binding pocket to ensure single unbinding events.

Rupture forces and loading rates were determined for each individual force curve. As for experiments, the loading rate was also determined from the slope of the force-time curve prior to rupture. We calculated the mean and standard deviation of the rupture forces for each loading rate.

**Brownian dynamics simulations**

To extract information on the underlying free energy landscape along the unbinding reaction coordinate, and in addition to the application of the simple Bell model and more sophisticated theoretical approaches (BSK (15), FNdY (37), DHS (14) and CHS (36)), numerical simulations of enforced unbinding with one-dimensional energy landscapes were carried out. To that end, the Smoluchowsky eq.

$$\partial_t p(x,t) = \nabla D \left[ \nabla - \beta \left( \nabla V(x,t) \right) \right] p(x,t),$$

was solved numerically for a time dependent potential $V(x,t) = V_0(x) - \dot{F}xt$, where $\beta = 1/k_B T$ and $\dot{F}$ is the applied loading rate. As underlying unbinding potential, a double barrier potential $V_0(x) = 2\Delta G_1(x)^2(1.5 - y(x))$ for $x \leqslant x_{\beta 1}$ was chosen, with barrier height $\Delta G_1$ of the first barrier. The function $y(x) = \left( \left( c_1 \frac{x}{x_{\beta 1}} + c_2 \right) \frac{x}{x_{\beta 1}} + \omega_1 \right) \frac{x}{x_{\beta 1}}$, with the constants $c_1 = \omega_1 + \omega_2 - 2$ and $c_2 = 3 - 2\omega_1 - \omega_2$, served to control the shape of the first barrier, particularly the curvatures $\omega_1$ and $\omega_2$ of the well and, respectively, that at the first barrier top. The above function also serves to control the rupture length $x_{\beta 1}$, here defined as the distance between first barrier top and minimum of the well of the bound state.

For $x_{\beta 1} < x < x_{\beta 2}$ the potential becomes

$$V_{x > x_{\beta 1}}(x) = V_0\left( x_{\beta 1} \right) + 2(\Delta G_2 - \Delta G_1)y(x)^2(1.5 - y(x))$$

with $y(x) = \frac{x - x_{\beta 1}}{x_{\beta 1}}$ and the potion of the second barrier $x_{\beta 2}$. For all $x > x_{\beta 2}$ the potential is set to $\Delta G_2$, the height of the second barrier.

Trajectories were generated starting from a Boltzmann ensemble $p_0(x)/\exp(V_0(x))$ within the well of the bound state. Positions were updated to first order according to the solution of the Smoluchowski equation for linear potential,

$$x_{new} := x_{old} - \frac{D\Delta t}{k_B T} \nabla V(x,t) + \frac{1}{\sqrt{2D\Delta t}} \xi(t),$$

with an integration time step of 0.5 ps, which ensured that less than 5% of the integration steps were larger than $0.05 x_{\beta 1}$. A Gaussian distributed (variance one) random force $\xi(t)$ was used. A diffusion constant $D = k_B T/\gamma = 4 \cdot 10^{-11} \, m^2/s$ was found to provide the best fit to the force spectrum.

To reduce computational effort and thus to facilitate the generation of trajectories even for the slowest loading rates, an appropriate biasing potential

$$V_{\text{bias}}(x,t) = a \left( \frac{x}{x_b} \right)^n + v(t)$$

was added whenever the (actual) barrier height exceeded $5 k_B$T, with

$$a = \frac{1}{nc^{n-1}}, c = \frac{nv}{n-1}$$

and $v$ adapted such that the barrier height of the resulting total potential $V(x,t) + V_{\text{bias}}(x,t)$ remained between 5 and 7 $k_B$T at all times. To compensate for the reduced barrier height due to the biasing potential and the increased barrier transition probability per integration step, the integration step size was dynamically rescaled by an acceleration factor $Z_{bias}(t) = Z_0(t)$, where $Z_0$ and $Z_{bias}$ are the partition functions of the unperturbed and, respectively, perturbed bound states

$$Z_0(t) = \int_{-\infty}^{x_b} \exp\left(-\beta V(x,t)\right)dx \text{ and } Z_{bias}(t) = \int_{-\infty}^{x_b} \exp\left(-\beta\left(V(x,t) + V_{\text{bias}}(x,t)\right)\right)dx.$$

For each of the loading rates for which experimental or MD simulation derived rupture forces were obtained (see **Fig. 2**), 1000 trajectories were generated and the resulting individual force at the point of barrier crossing were averaged. The fitting parameters $\Delta G_1$, $\Delta G_2$, $x_{\beta 1}$, $x_{\beta 2}$, $\omega_1$, and $\omega_2$ were varied using simple line-scanning until the $\chi^2$ (weighted by the standard error of the mean) relative to the 37 experimental/simulation averaged rupture forces was minimal. Uncertainties of the six fitting parameters (given in the caption of **Fig. S7**) were estimated conservatively via non-parametric bootstrapping with 80 replica data sets, each of which was generated by randomly drawing 37 from the 37 rupture forces (allowing for multiple draws of the same data point). Parametric bootstrapping using a Gaussian error model and the standard errors of the mean rupture forces yielded slightly smaller uncertainties.

**Center of Mass distributions**

To calculate the distributions of the center of mass (COM) distance of each MD trajectory, we calculated the distance between the center of mass of the binding pocket forming residues (L25, S27, Y43, S45, V47, G48, A50, W79, R84, A86, S88, T90, W92, W108, L110, D128) and the center of mass of biotin. This was done starting shortly before rupture and ending when the distance between biotin and the binding pocket was larger 4 nm.

As we do not see any binding patterns further away than 2 nm we reduced the plotted histograms to a maximal distance of 2 nm. The COM distances were then binned into 200 equally spaced bins and normalized by the total amount of data points.

**MD intermediate states and transition plots**

To determine the intermediate states, we split the COM distance distributions according to the dominant peaks, such that each peak represents a different intermediate state. We then calculated the transition rates by counting the transitions from state $s_i$ to state $s_j$ and subtracted the amount of back-transitions ($s_j$ to $s_i$) for each combination, resulting in a net transition of -1, 0 or 1. Finally each rate was averaged for each velocity.

The probabilities for being in an intermediate state where calculated by counting the total time that an intermediate state is visited normalized by the total time spent in all intermediate states. The time spent in the ground state and the unbound state were not taken into account to provide comparability, as the time spent in either of them is arbitrary and depends only on the chosen starting- and endpoint of the rupture event.

**Principal component analysis of loop 3-4**

In order to understand the different unbinding pathways, we analyzed the motion of loop 3-4 by performing a principal component analysis (PCA) on the backbone atoms of residues 44 – 54. Therefore, a representative simulation was used to calculate the characteristic Eigenvectors of the loop 3-4 motion going from a closed conformation (negative values) to an open conformation (positive values). All other simulations were projected onto the first Eigenvector and analyzed depending on the applied loading rate regime. The time resolved projections were finally binned along the COM distance and the average projection on the first Eigenvector for each loading rate regime was calculated (SI Fig. S10).

**Acknowledgements**

Discussion of FR and HG with Prof. Klaus Schulten were inspiring for this work. FR and SS thank Prof. Toshio Ando for generously providing AC7 cantilevers. This work was supported in part by the Agence National de la Recherche Grant Nos. ANR-15-CE11-0007 (BioHSFS) and ANR-11-LABX-0054 (Labex INFORM), COST Action TD1002-10006, and by the Deutsche Forschungsgemeinschaft (DFG) Grant No. SFB 755.A5.

# Supplementary Information for

## Heterogeneous and rate-dependent streptavidin-biotin unbinding revealed by high-speed force spectroscopy and atomistic simulations

Felix Rico[1‡*], Andreas Russek[2‡], Laura González[3], Helmut Grubmüller[2*], and Simon Scheuring[4,5*]

[1] LAI U1067, Aix Marseille Univ, INSERM, CNRS, 163 avenue de Luminy, 13009 Marseille, France

[2] Department of Theoretical and Computational Biophysics, Am Faßberg 11, Göttingen, Germany

[3] Bioelectronics Group, Department of Electronics, Universitat de Barcelona, c/ Marti Franques 1, 08028 Barcelona, Spain

[4] Department of Anesthesiology, Weill Cornell Medical College, 1300 York Ave, New York, NY 10065, USA

[5] Departments of Physiology and Biophysics, Weill Cornell Medical College, 1300 York Ave, New York, NY 10065, USA

[*]Email:  felix.rico@inserm.fr, sis2019@med.cornell.edu, hgrubmu@gwdg.de

**This PDF file includes:**

Supplementary text
Figures S1 – S10
Movies S1 – S6
References for SI reference citations

**Supplementary Information Text**

**Previous single molecule force spectroscopy of the streptavidin/biotin complex.** The complex formed by streptavidin (SA) and the small molecule biotin (b, vitamin H) is one of the strongest non-covalent bonds known in nature. Monomeric streptavidin forms the biotin-binding pocket with an eight-stranded, antiparallel beta-barrel capped by loop 3-4. In the native, tetrameric SA, loop 7-8 from an adjacent monomer contributes to closing the binding pocket and was shown important to preserve the binding properties of the complex (59). Biotin binds by forming an intricate network of hydrogen bonds with polar residues of both the beta-barrel and loop 3-4 (4, 5). Due to its high affinity ($K_D \sim 10^{-13}$ M) and long lifetime ($k_{off} \sim 10^{-6}$ s$^{-1}$, $\sim 0.1$ day$^{-1}$) as measured from bulk, equilibrium measurements (6-9), the SA-b complex is extensively used in molecular biology for protein labeling and purification. SA-b is also widely used in biotechnology and single molecule measurements to immobilize proteins, DNA and RNA molecules on surfaces as it supports large pulling forces (60). The forced disruption of the streptavidin-biotin complex by atomic force microscopy (AFM) and other techniques established the basis of single molecule biomechanics (8-12, 14, 16-21, 28, 61-63). However, single molecule experiments on the SA-b complex often estimated the intrinsic unbinding rate to be orders of magnitude faster than off-rates measured from bulk-equilibrium experiments ($\sim 10^{-6}$ s$^{-1}$), and data diverge, although this divergence seems to disappear if low binding frequency (highe probability of probing single bonds) is assured (49). Nevertheless, also bulk experiments have reported dissimilar intrinsic unbinding rates (6, 7). All atom steered molecular dynamics simulations (SMDS) of the forced unbinding of SA-b were pioneer in the field and have provided mechanistic descriptions of the unbinding process (26, 27). However, due to limited computational power for SMDS, the number of studies assessing the dynamic nature of the bond are still rare and were performed either on monomeric SA or at pulling velocities several orders of magnitude faster than experiments, precluding direct comparison (26-28, 61, 64-67). Thus, knowledge of the dynamic nature of the SA-b bond and its (un)binding process is still limited, i.e. it is still unknown how this non-covalent bond outlives days.

**Bimodal distributions.** Bimodal distributions in force histograms have been attributed to simultaneous double bond rupture events and cooperative effects at the high forces reached at fast loading rates (11, 16, 63, 68). Using cantilevers with shorter time response minimized the occurrence of bimodal distributions (**SI**), probably because they allowed easier discrimination between curves with single and double rupture events even at the highest velocities.

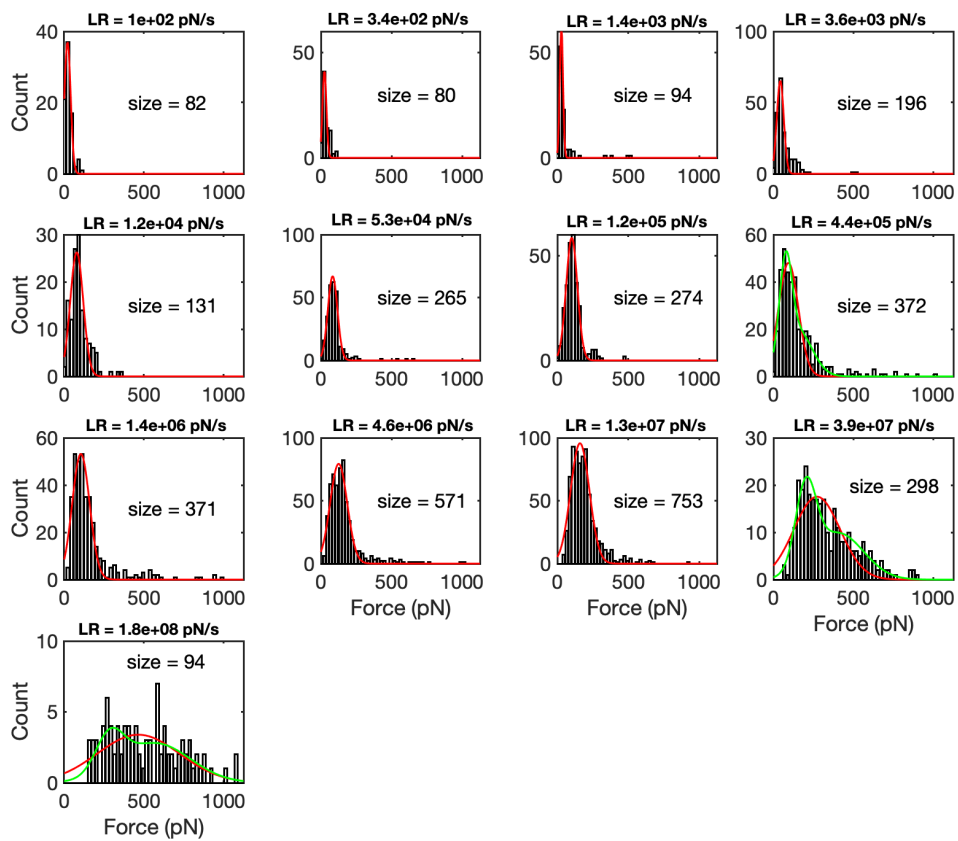

**Fig. S1.** Rupture force distributions using AC10 cantilevers at different loading rates (LR) indicating the number of analyzed events per plot. Solid lines are best fits to a single (red) and double (green) gaussian distributions. When a double distribution was used, the most probable force was determined from the first peak.

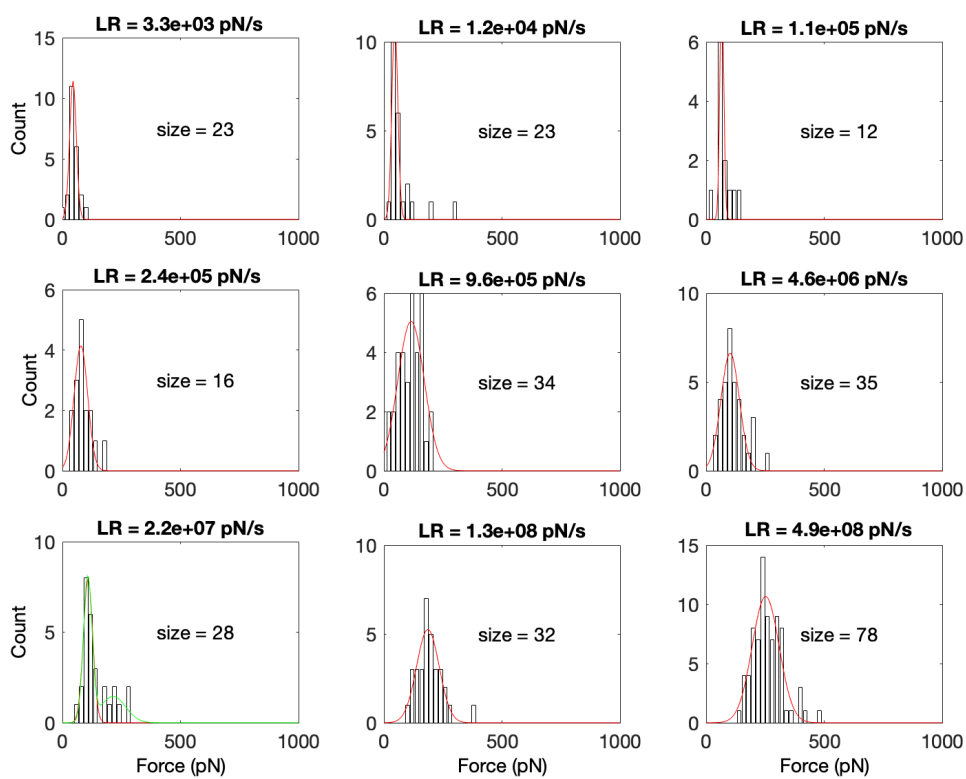

**Fig. S2.** Rupture force distributions using AC7 cantilevers at different loading rates (LR) indicating the number of analyzed events per plot. Solid lines are best fits to a single (red) and double (green) gaussian distributions. When a double distribution was used, the most probable force was determined from the first peak.

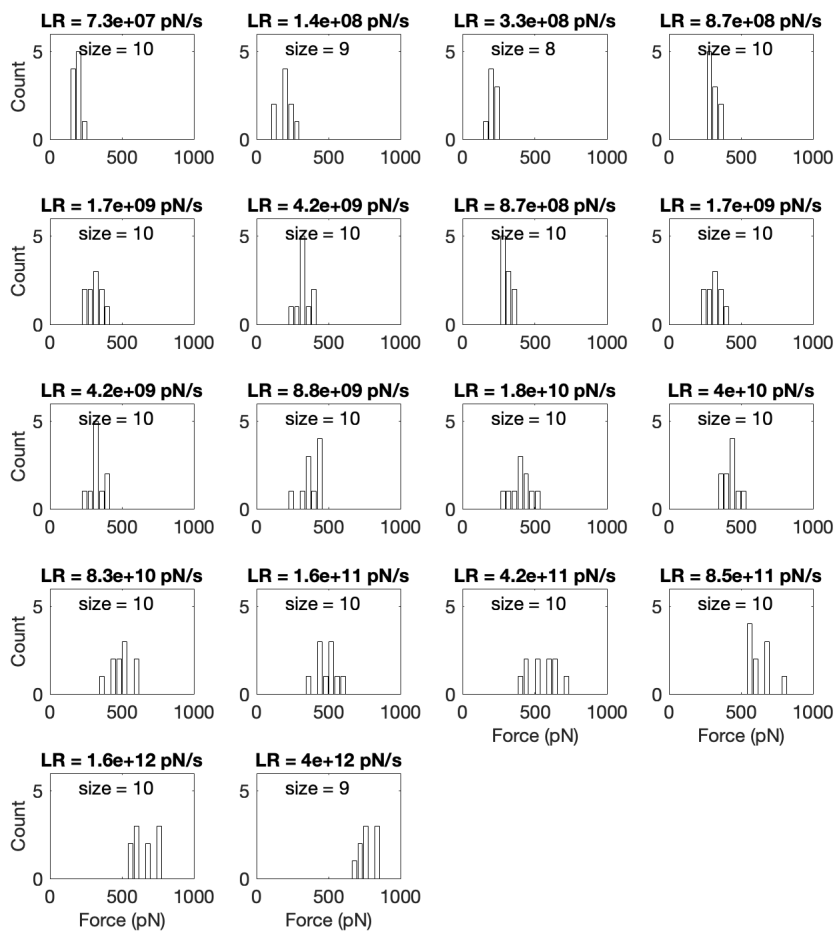

**Fig S3.** Rupture force distributions from tetrameric SA/b MD simulation trajectories at different loading rates (LR) indicating the number of analyzed events per plot.

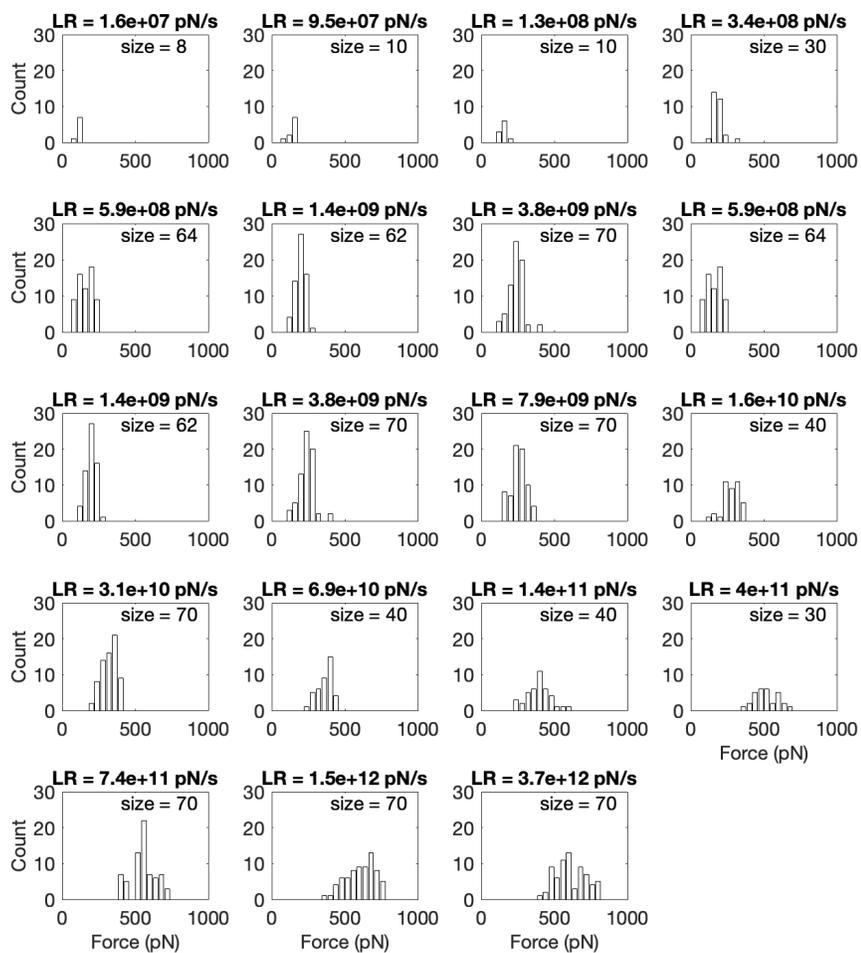

**Fig S4.** Rupture force distributions from monomeric SA/b MD simulation trajectories at different loading rates (LR) indicating the number of analyzed events per plot.

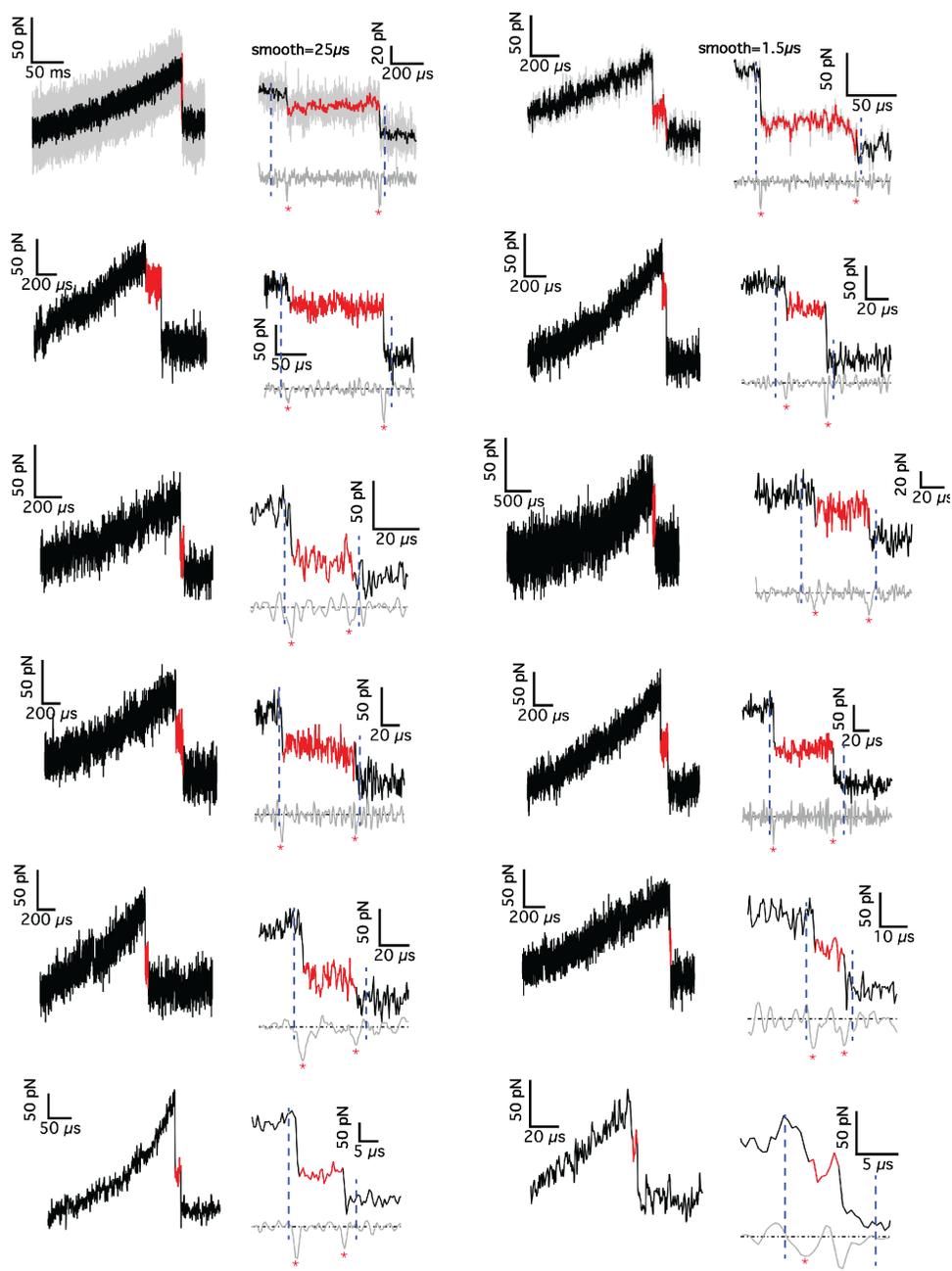

**Fig S5.** Examples of rupture events (left) showing the zoomed transient binding (right). Raw curves are represented as solid black lines (in grey if smoothed) with transient binding events shown in red. The numerical derivate is shown in grey below the zoomed region (smoothed with a window between 1.5 µs and 25 µs according to pulling velocity and sampling rate). Vertical, dashed blue lines delimit the force drop region analyzed in search of transient binding events. The two most prominent minima within this region are shown with asterisks and were used to determine the lifetime and average force of the transient binding events.

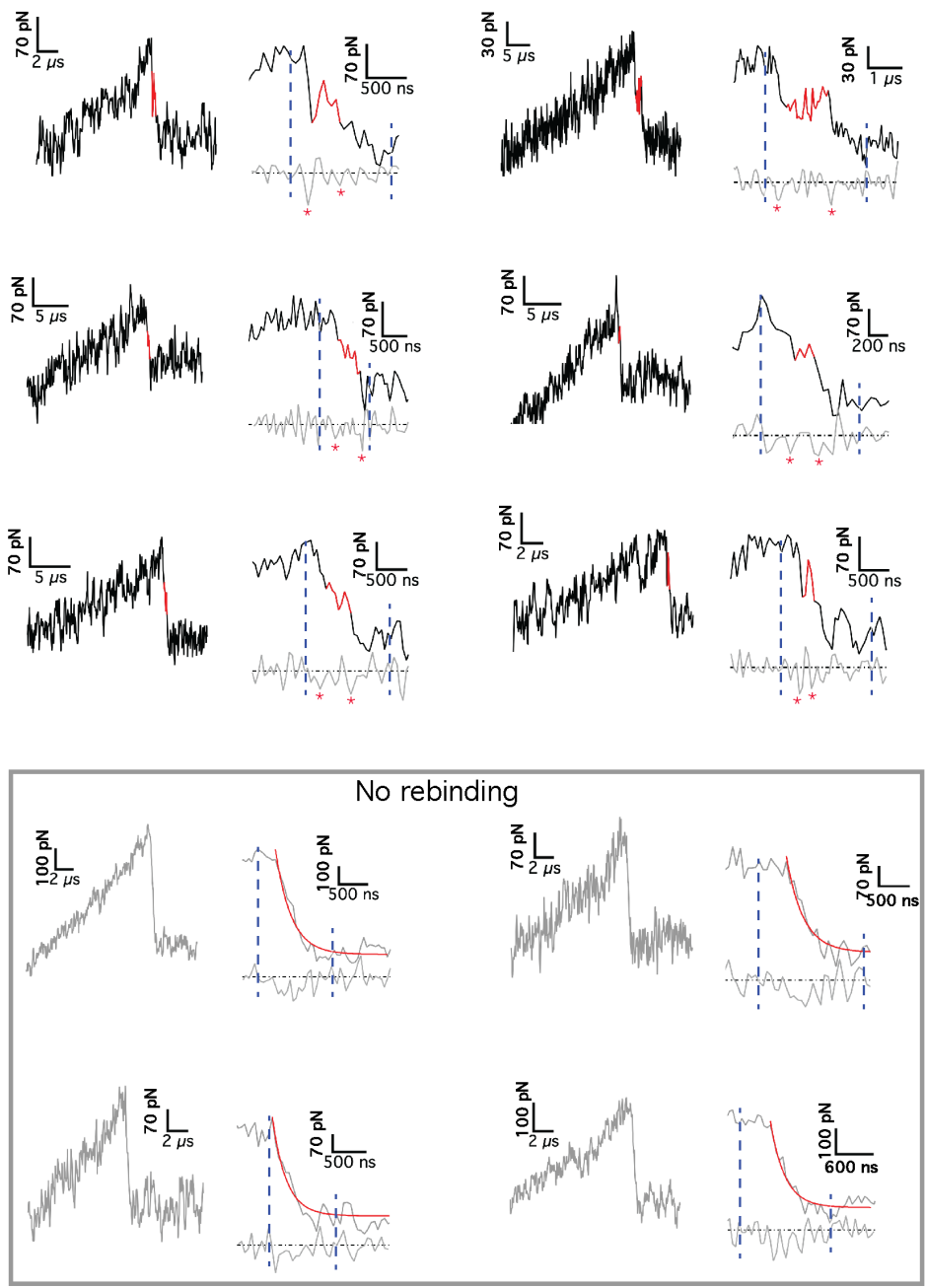

**Fig S6.** Top six curves: Examples of rupture events (left) showing the zoomed transient binding (right). Raw curves are represented as solid black lines with transient binding events shown in red. The numerical derivate is shown in grey below the zoomed region (no smoothing used. Vertical, dashed blue lines delimit the analyzed region in search of intermediate binding events. The two most prominent minima within this region are shown with asterisks and were used to determine the lifetime and average force of the transient binding event. Four bottom curves: Examples of rupture events without transient binding events (grey solid lines). The red line represents an exponential fit with decay time ranging from 0.22 to 0.32 μs.

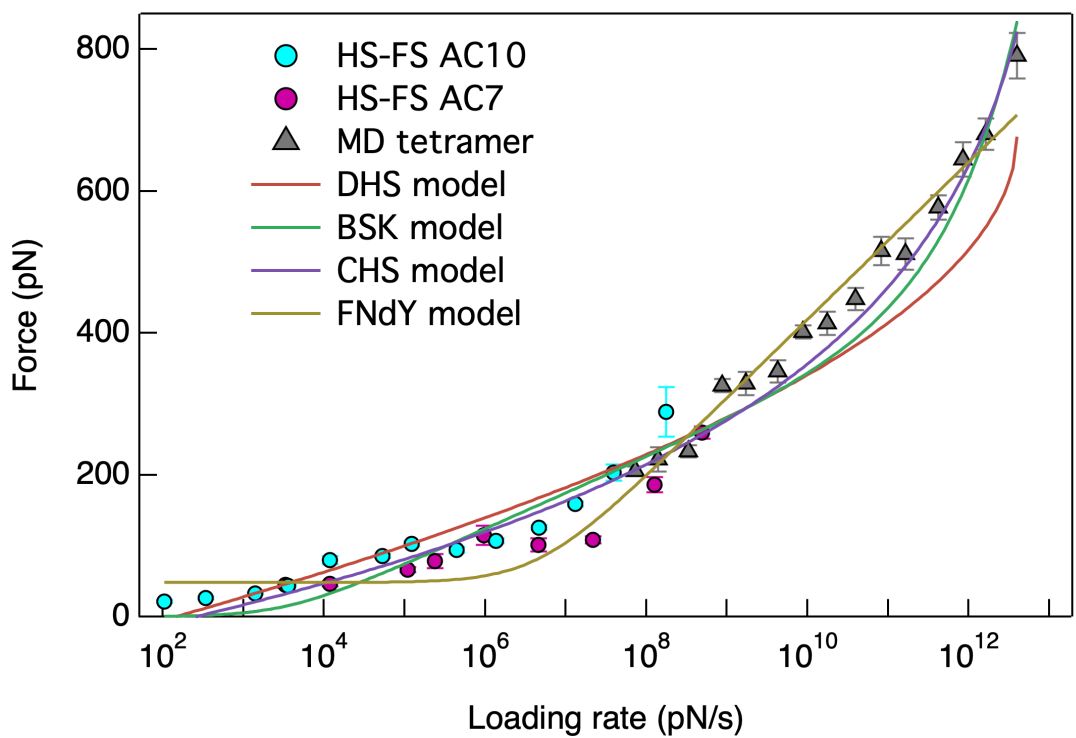

**Fig S7.** Experimental and MD simulations dynamic force spectra of the SA/b interaction. Solid lines are the best fits of the BSK (15), FNdY (37), DHS (14) and CHS (36) models to the combined dynamic force spectrum. Analytical models were fitted to data using nonlinear least squares weighted with the standard error of the mean. The fitted parameters for each model with error estimations (standard deviation) are reported below. The reported $\chi^2$ is renormalized by the average number of measurements per data point.

**DHS**

$\chi^2$ = 10.4
$k^0$ = 5.7 ± 2.9 s⁻¹
$x_\beta$ = 0.29 ± 0.02 nm
$\Delta G$ = 24.0 ± 0.1 $k_B T$

**BSK**

$\chi^2$ = 7.7
$D$ = (3.9 ± 1.1) x 10⁷ nm²/s
$x_\beta$ = 0.18 ± 0.01 nm
$\Delta G$ = 20.4 ± 0.1 $k_B T$

**FNdY**

$\chi^2$ = 10.6
$F_{eq}$ = 48 ± 6 pN
$x_\beta$ = 0.08 ± 0.01 nm
$k^0$ = 19646 ± 8490 s⁻¹

**CHS**

$\chi^2$ = 4.9
$k^0$ = 0.02 ± 0.01 s⁻¹
$x_\beta$ = 0.34 ± 0.03 nm
$\Delta G$ = 25.0 ± 0.1 $k_B T$
$\mu$ = (2.7 ± 17) x 10⁻⁵

**BD**

$\chi^2$ = 4.8
$\Delta G_1$ = 16.9 ± 0.2 $k_B T$
$\Delta G_2$ = 20.7 ± 0.7 $k_B T$
$x_{\beta 1}$ = 0.19 ± 0.02 nm
$x_{\beta 2}$ = 0.44 ± 0.07 nm
$\omega_1$ = 0.9 ± 0.4
$\omega_2$ = 1.7 ± 0.7
$D$ = 4.0 x 10⁷ nm²/s

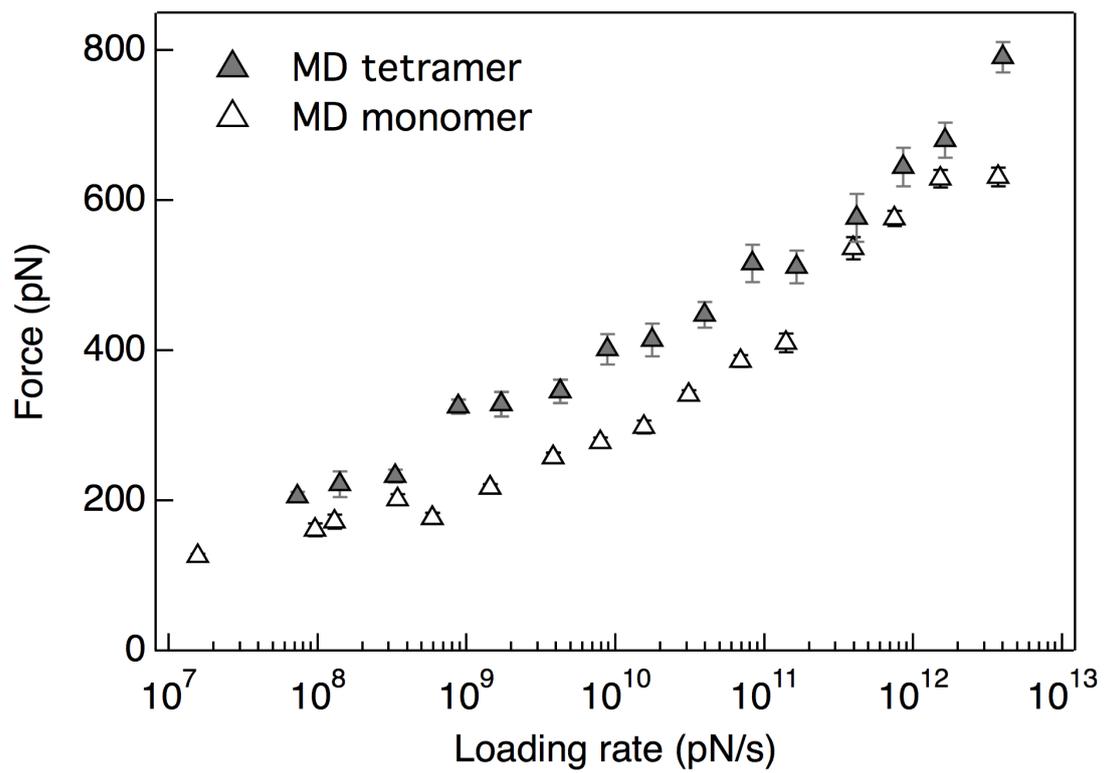

**Fig. S8.** Comparison between monomeric and tetrameric streptavidin-biotin MD force spectra.

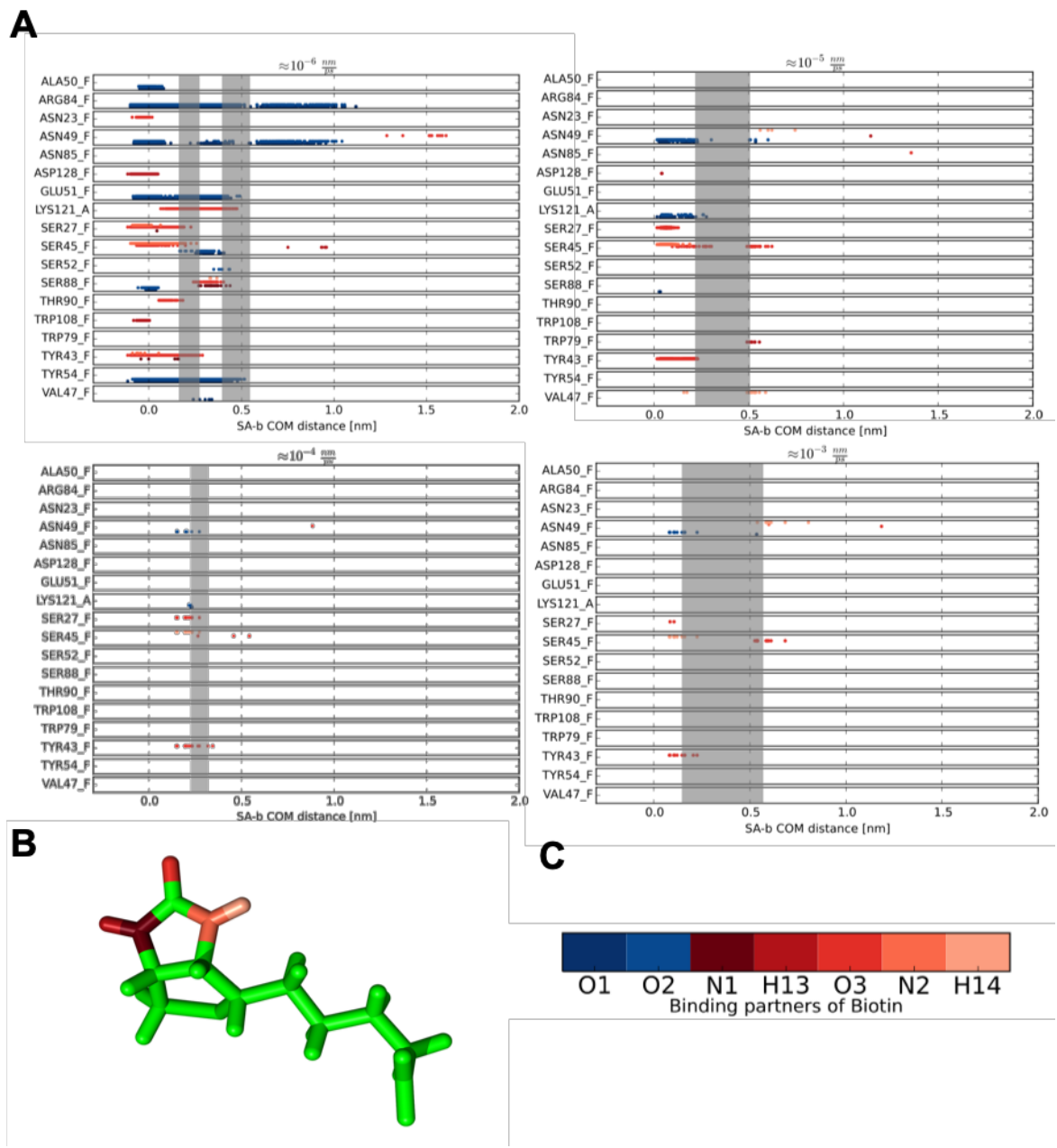

**Fig. S9.** Examples of simultaneous bond rupture of H-bonds with subsequent building of new binding patterns (grey background) at the 4 different velocity regimes named in the titles. For easier comparison, each subplot shows all interacting amino acids, even if no H-bond was formed. The existence of H-bonds between the named amino acids and biotin is shown as colored dots. F stands for the SA monomer out of which biotin is pulled and A (see Lys121) for an adjacent SA monomer. The color of the dots refers to the interaction partner of biotin shown in B as stick representation. Red colors stand for interactions with the head group of biotin as shown in B. The blue dots are H-bonds formed with the tail of biotin. C Color bar showing the used color-code in A.

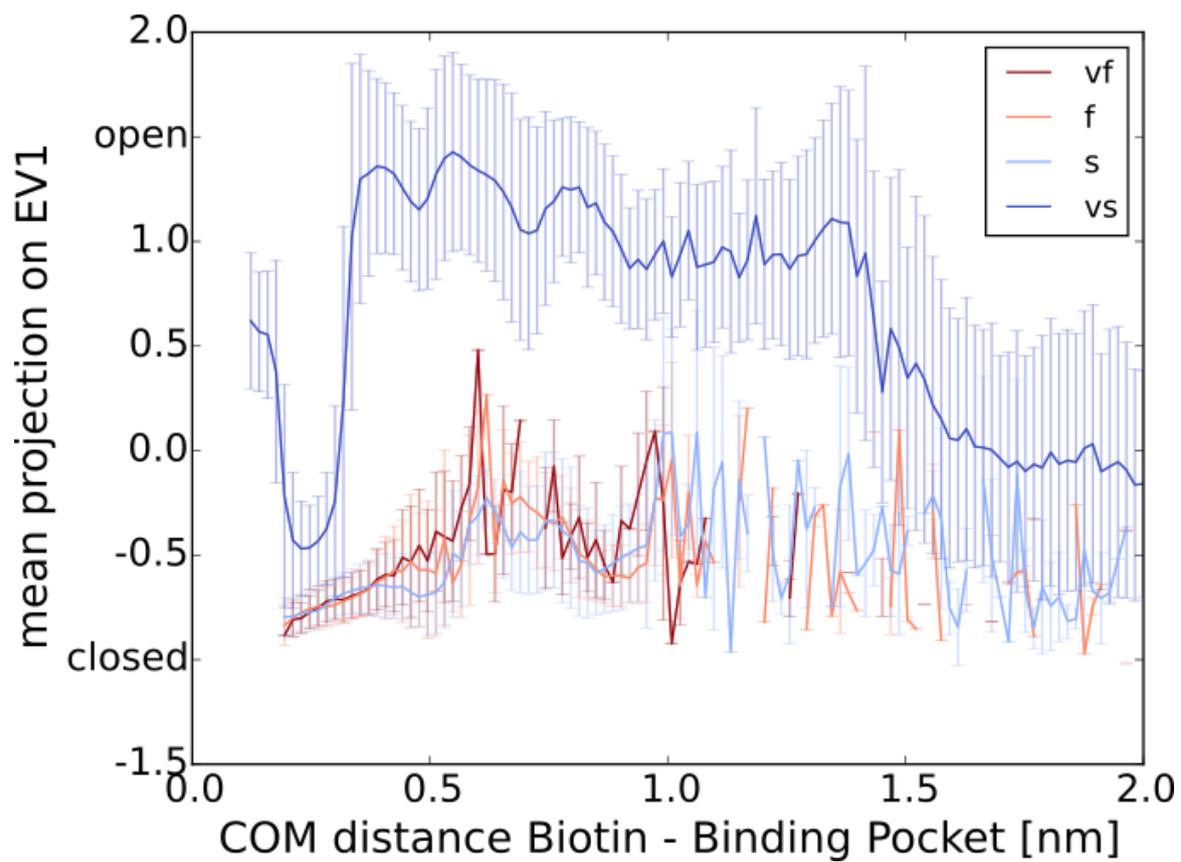

**Fig. S10.** Projection of all trajectories onto the first eigenvector of a PCA of the loop3-4 motion plotted over the COM distance between the binding pocket and biotin. Different colors denote different pulling velocity regimes ranging from very fast (vr, red) to very slow (vs, blue).

**Movies**

**Video S1.** Representative movie of an unbinding trajectory from MD simulations of tetrameric SA/b at 1 m/s pulling velocity.

**Video S2.** Representative movie of an unbinding trajectory from MD simulations of tetrameric SA/b at 0.2 m/s pulling velocity.

**Video S3.** Representative movie of an unbinding trajectory from MD simulations of tetrameric SA/b at 0.01 m/s pulling velocity.

**Video S4.** Representative movie of an unbinding trajectory from MD simulations of monomeric SA/b at 1 m/s pulling velocity.

**Video S5.** Representative movie of an unbinding trajectory from MD simulations of monomeric SA/b at 0.1 m/s pulling velocity.

**Video S6.** Representative movie of an unbinding trajectory from MD simulations of monomeric SA/b at 0.01 m/s pulling velocity.